\documentclass{aa}
\newcommand{\h}{h_{50}}
\newcommand{\BF}{}

\begin{document}
%\begin{titlepage}
%
\thesaurus{ 
	    02.02.1;      %black hole physics
	    11.09.3;      %intergalactic medium
	    11.10.1;      % galaxies: jets
	    11.03.4: Coma;%Galaxies: clusters: individual: Coma
	    12.12.1}      %(Cosmology:) large-scale structure of the Universe

\title{Black Hole Energy Release to the Gaseous Universe}
\author{Torsten A. En{\ss}lin\inst{1} \and Yiping Wang\inst{1,2,3}
\and Biman B. Nath\inst{4} \and Peter L. Biermann\inst{1}}
\institute{Max-Planck-Institut f\"{u}r Radioastronomie, Auf dem 
H\"{u}gel 69, D-53121 Bonn, Germany \and Purple Mountain Observatory,
Academia Sinica, Nanjing 210008, China \and Bergische Universit\"at Wuppertal,
D-42097 Wuppertal, Germany \and Raman Research Institute, Bangalore
560080, India}
\offprints{T.A. En{\ss}lin, ensslin@mpifr-bonn.mpg.de}
\date{Received ??? , Accepted ???}  
 \maketitle\markboth{En{\ss}lin et al.: Black Hole Energy Release to
the Gaseous Universe}{En{\ss}lin et al.: Black Hole Energy Release to
the Gaseous Universe}
\begin{abstract}
We estimate the energy release of black hole formation to the intra-cluster
medium of the Coma cluster and find that this is comparable to the present
day energy content. Therefore an energetic and maybe hydrostatic influence
is possible. Our calculations rely on the assumption of an universal black
hole to galaxy mass ratio (more exactly: spheroidal mass component of the
stellar population), for which there is growing evidence. On a cosmological
scale, there is also an energy release of black hole formation comparable to
what is expected to be present within the thermal gas, caused by the process
of structure formation. This indicates an important dynamical influence,
neglected by present day structure formation simulations. This estimate of
cosmological black hole energy release is independent of the black hole to
galaxy mass ratio, but consistent with its value.
\end{abstract}
%\end{titlepage}

\section{Introduction}
There is growing evidence for the existence of black holes (BHs) in all
galaxies, being remnants of the strong quasar activity in the young Universe
(Lynden-Bell 1969; Rees 1989; Haehnelt \& Rees 1993; van der Marel 1997;
Ford et al. 1997; Hasinger 1998; and references
therein)\nocite{lynden-bell69, rees89, haehnelt93, vdMarel97,
hasinger98}. Estimates of the ratio of the BH mass to that of the spheroidal
component of the stellar population of the host galaxy seem to converge to
$\eta_{\rm bh}\approx 0.002 - 0.006$ (Kormendy \& Richstone 1995; Faber et
al. 1997; Magorrian et al. 1997; Silk \& Rees 1998; Wang \& Biermann
1998\nocite{wang98}). This allows to estimate the total BH mass within a
population of galaxies. The former BH growth was accompanied by an output of
large amounts of energy by radiation, heating of the ambient medium and
outflows of relativistic particles, which can influence the environment of
the host galaxies energetically and dynamically. In this Letter we present a
comparison of the BH energy release and the thermal energy of the quasar
environment in order to demonstrate the possible importance of this
influence. The efficiency of energy dissipation of the accreting matter to a
Schwarzschild BH is $\varepsilon\approx 0.06$, whereas to a maximal rotating
BH $\varepsilon\approx 0.3$ (Thorne 1974; Laor \& Netzer
1989)\nocite{thorn74,laor89}.  We assume $\varepsilon_{\rm light}\approx
0.1$ to be the mass-to-light conversion rate, and $\varepsilon_{\rm
th+nth}\approx 0.1$ for the production efficiency of thermal and nonthermal
energy release in relativistic particles and magnetic fields. Therefore we
use an intermediate value of $\varepsilon \approx 0.2$. We adopt a
conservative value of $\eta_{\rm bh}\approx 0.002$, since we like to include
the fraction of galaxies containing massive BHs into this number. If
$\eta_{\rm bh}\approx 0.006$ our estimate of energy release to the Coma
cluster is still correct if only one third of all galaxies contain massive
BHs. But the cosmological estimate is not affected, since it depends only on
the ratio $\varepsilon_{\rm th+nth}/\varepsilon_{\rm light}$ and the
observed quasar light.  Further, we use $H_{\rm o} = 50\,\h\, {\rm km\,
s^{-1}\, Mpc^{-1}}$, $\Omega_{\rm o} = 1$, $\Lambda = 0$, and indicate the
scaling of all quantities with the Hubble constant.

\section{The Coma Cluster}
First, we estimate the thermal energy content of the Coma cluster.  The
electron density of the intra-cluster medium of the Coma cluster can be
described by a $\beta$-model: $n_{\rm e}(r)= n_{\rm e,o} [ 1+ (r/r_{\rm
core})^2 ]^{-3\beta_{\rm e} /2}$, where $n_{\rm e,o} = 3\cdot
10^{-3}\,\h^{1/2}\, {\rm cm^{-3}}$ is the central electron density, $r_{\rm
core}= 400\,\h^{-1}\, {\rm kpc}$ the core radius, and $\beta_{\rm e}= 0.75$
gives the slope (Briel et al. 1992)\nocite{briel92}. The gas extends to a
radius of $R_{\rm cluster}\approx 5\,\h^{-1}\,{\rm Mpc}$, where an accretion
shock of the infalling matter marks the cluster boundary (En{\ss}lin et
al. 1998)\nocite{ensslin98b}. The integrated gas mass within this radius is
$4.2\cdot 10^{14}\,\h^{-5/2}\, {\rm M_\odot}$. The central temperature of the
cluster is $8.2\, {\rm keV}$ (Briel et al. 1992)\nocite{briel92}, but there is
evidence for a temperature decrease from the center to the boundary by a
factor of two (Fusco-Femiano \& Hughes 1994\nocite{fusco-femiano94}; Honda et
al.  1996\nocite{honda96}), which seems to be typical for clusters (Markevitch
et al. 1997)\nocite{markevitch98}. We therefore describe the temperature with
a $\beta$-profile with the same core radius, but with $\beta_{kT}= 0.09$,
leading to a temperature drop by two within $5\,\h^{-1}\, $Mpc. The resulting
thermal energy content within $R_{\rm cluster}$ is $1.3\cdot
10^{64}\,\h^{-5/2}\, {\rm erg}$.

This has to be compared with the energy release of BH formation.
Since the galaxies in Coma are mainly ellipticals, only their total mass has
to be estimated and multiplied with $\eta_{\rm bh}$ in order to get the mass
of BHs in Coma. The mass of the ellipticals can be derived from their
luminosity function and their mass to light ratio.  The R-band luminosity
function of early type galaxies in the central $700\, {\rm arcmin^2}$ of Coma
is given in Secker \& Harris (1996)\nocite{secker96}.  The complete
luminosity function can be estimated by assuming these galaxies to be within
a cylindrical volume, and correcting to the whole cluster volume, with the
help of the spatial distribution of galaxies. The radial distribution of
galaxies within Coma is given by $n_{\rm gal} \sim (1 + (r/r_{\rm gal})^2)^{
-\alpha_{\rm gal}}$, with $\alpha_{\rm gal}=0.8$ and $r_{\rm gal} =
160\,\h^{-1}\, {\rm kpc}$ (Girardi et al. 1995)\nocite{girardi95}. Including
into this the luminosities of NGC 4874 and NGC 4889 (Strom \& Strom
1978)\nocite{strom78}, which were not included into this luminosity
function, and applying a correction of V-R = 0.85, which is typical for
ellipticals (Tinsley \& Gunn 1976\nocite{tinsley76}), we get the V-band
luminosity function. The galactic mass can now be integrated with the help of
the mass to (V-band) light ratio given by Magorrian et
al. (1997)\nocite{magorrian98}. It is $M_{\rm gal}= 2.9\cdot 10^{13}\,\h^{-1}
\, {\rm M_\odot}$, giving a total BH mass of $M_{\rm bh} = 5.8\cdot
10^{10}\,\h^{-1} \, (\eta_{\rm bh}/0.002)\, {\rm M_\odot}$.  The
gravitational energy, which was freed during the BH formation, is
therefore $E_{\rm bh} =\varepsilon M_{\rm bh}c^2 = 2.1\cdot 10^{64}\,\h^{-1}
\, (\varepsilon/0.2)\, (\eta_{\rm bh}/0.002)\, {\rm erg}$.  The thermal plus
nonthermal energy release (one half of the total) is therefore comparable to
the present day thermal energy content of Coma within a $5\,\h^{-1}\, $Mpc
radius. 

A different approach to the nonthermal energy release of BH formation to
clusters of galaxies was taken by En{\ss}lin et al. (1997) by an estimate of
the jet-power of radio galaxies, integrated over cosmological epochs.  A
comparison with the typical thermal energy content of cluster showed, that
both numbers could be comparable on cluster scale, if a considerable amount of
energy is ejected by the not radio-emitting and therefore invisible
relativistic protons. Our above estimate of the nonthermal energy release is
completely independent of this calculation, and therefore demands such
energetic protons, having energy densities much higher than the electrons
within the relativistic plasma flowing out of radio galaxies, consistent with
theories of the origin of the observed ultra-high-energy cosmic rays (Rachen
\& Biermann 1993; Biermann 1997)\nocite{rachen93} and recently discovered 
TeV $\gamma$-rays from blazars (Mannheim 1998)\nocite{mannheim98}.

Since also the nonthermal energy should be stored in the intra-cluster
medium for at least a Hubble time, we expect a large nonthermal energy
content of Coma and similar clusters. Discussions of the storage of
relativistic particles in clusters can be found in V\"olk et al.
(1996)\nocite{voelk96} (who estimate the supernovae energy release into the
Perseus cluster), En{\ss}lin et al. (1997)\nocite{ensslin97}, and Berezinsky
et al. (1997)\nocite{berezinsky97}. We note, that details of the injection
of relativistic plasma from quasars into the intra-cluster medium and also
the later evolution of this plasma need further investigations. E.g. in
speculative scenarios of dwarf galaxy formation induced by violent expansion
of quasar outflows in the early Universe large fractions of the released
energy are transformed to heat in the ambient medium by shocks and adiabatic
compression (Natarajan et al. 1997; Silk \& Rees 1998; and
references therein)\nocite{natarajan97,silk98}. The question if the
nonthermal plasma influences the hydrostatics of the cluster, and should be
taken into account into (uncertainties of) X-ray mass determinations, can
not be answered yet, since this depends on the amount and also crucial on
the spatial distribution of the nonthermal phases.

At the location of the radio outflow from Perseus A an X-ray hole is present
in the thermal emission of the Perseus cluster (B\"ohringer et al. 1993),
demonstrating that the radio plasma rather displaces the thermal gas, driven
in this case with a power of probably more than $10^{46}\, {\rm erg\,
s^{-1}}$ (Heinz et al. 1998), than mixes with it.  Thus the radio plasma
might remain within the intra-cluster medium, invisible since the electrons
cool down in a cosmologically short time, and might be reactivated when a
cluster merger event reaccelerates the electron population, as it could be
the case for the $3\,\h^{-1}\, $Mpc sized, diffuse radio halo of Coma (Deiss
et al.~1997\nocite{deiss97}; and references therein). Also the peripherally
located cluster radio relics (e.g. 1253+275 in Coma) can be understood in
terms of old reactivated radio plasma, having passed through a cluster
accretion- or merger shock wave (En{\ss}lin et al. 1998).  Evidence for the
existence for a necessary, old, but large relativistic electron population
within the intra-cluster medium, having energies below what is visible in
the radio, can be seen in the recently detected EUV excess of the Coma
cluster (Lieu et al. 1996), if this is explained by inverse-Compton
scattered microwave background photons (Hwang 1997; En{\ss}lin \& Biermann
1998; Sarazin \& Lieu 1998)\nocite{hwang97,ensslin98a,sarazin98}.

\section{The Universe}
It is also possible to compare the BH formation energy release to that of
the thermal gas on a cosmological scale. The integrated energy density in
quasar light emitted during (and estimated for) the epoch of quasar activity
is $(0.9 - 1.3)\cdot 10^{-15}\, {\rm erg\, cm^{-3}}$ (Chokshi \& Turner
1992; an earlier, more conservative estimate can be found in Soltan
1982)\nocite{chokshi92, soltan82}. This number is given per comoving volume,
and it is independent of cosmology. Chokshi \& Turner (1992) assumed an
efficiency of light production of $\varepsilon_{\rm light}\approx 0.1$, and
got a present day BH mass density of $(1.4 - 2.2)\cdot 10^{5}\, {\rm
M_\odot\, Mpc^{-3}}$, consistent with the above used BH to spheroidal mass
ratio $\eta_{\rm bh}$, even if all galaxies contain BHs (Faber et al.
1997)\nocite{faber97}.

A large fraction of the energy output is injected into clusters,
roughly $f_{\rm cl} = 0.3 - 0.5$ (En{\ss}lin et
al. 1997)\nocite{ensslin97}, which contain less than 10\% of all
baryons, and therefore get more energy per baryon from BHs than the
gas outside clusters. Distributing 60\% of the BH energy release onto
the 90\% of the gas which is outside clusters gives an average
temperature of $kT(z_{\rm inj})$ $\leq (0.9-1.3)\, {\rm keV}$ $
(\varepsilon_{\rm th+nth} /\varepsilon_{\rm light})\,$ $((1-f_{\rm
cl})/0.6)$ $(\Omega_{\rm b}\h^{2} /0.05\,)^{-1}$ at the redshift of
injection.  This estimate assumes complete dissipation of the
nonthermal energy in magnetic fields and relativistic particles. This
is not realistic, but gives us upper limits to the heating and
Comptonization parameter. The true thermal heating might only be some
fraction of this number.

In a homogeneous universe comoving energy densities have to be
redshifted by $(1+z_{\rm inj})^{-\alpha}$ due to adiabatic expansion
losses, with $\alpha = 3 (\gamma -1)$ and $\gamma$ the adiabatic
index.  This correction has to be applied for the quasar light, which
has $\gamma_{\rm light} =4/3$ and therefore $\alpha_{\rm light} = 1$.
The average injection redshift of the luminosity function used in
Chokshi \& Turner (1992) is $z_{\rm inj} = 2.34$, giving a present day
quasar light density of $(0.3-0.4)\cdot 10^{-15}\, {\rm erg\,
cm^{-3}}$.  The redshift correction for BH thermal energy release
should be different, due to $\gamma_{\rm gas} =5/3$, but also since
this heat is injected into the environment of the quasar host
galaxies, which is the inter-galactic medium of the filaments and
sheets of the large scale structure and therefore partly decoupled
from the Hubble flow. We assume that the density in sheets and
filaments decreases with $(1+z)^d$ only due to the action of the
Hubble flow in the direction of these structures of dimension $d$: we
adopt $d=2$ for sheets and $d=1$ for filaments. Since most galaxies
are located in filaments, we use $d=1$, but give the scaling for
different values.  This is conservative, since these structures are
forming, and the ongoing infall of matter could lead to compression,
whereas the Universe still expands.
Thus, $\alpha_{\rm gas} = d(\gamma_{\rm gas} -1) = 2/3$ gives a
present day average heat (from BHs) outside clusters of
\begin{eqnarray}
kT(z=0) &\leq &(0.4-0.6)\,\, {\rm keV}\, 
\frac{\varepsilon_{\rm th+nth}}{\varepsilon_{\rm light}}\,
\frac{1-f_{\rm cl}}{0.6}\,
\nonumber\\ & &
\frac{(1+z_{\rm inj})^{-\alpha}}{3.34^{-2/3}}\,
\left(\frac{\Omega_{\rm b}\h^{2}}{0.05} \right)^{-1},\nonumber
\end{eqnarray}
(independent of cosmology) again assuming complete dissipation of nonthermal
phases (therefore $\leq$).  The corresponding Comptonization parameter $y \leq
(0.7-1.2)\cdot 10^{-5}\h^{-1}$, estimated for this adiabatic cooling history
and an Einstein-de-Sitter Universe (using $\Omega_{\rm o} =0$ increases $y$ by
only 50\%), does not violate the present day upper limit of $y < 1.5\cdot
10^{-5}$ (Fixsen et al. 1996)\nocite{fixsen96}.  Additional to the dependence
on $\h$, $y$ scales with $(1+z_{\rm inj})^{3/2}$ $ (3/2 + \alpha_{\rm
gas})^{-1}\, (\varepsilon_{\rm th+nth} /\varepsilon_{\rm light})$ $(1-f_{\rm
cl})$ in an Einstein-de-Sitter Universe, and $(1+z_{\rm inj})^{2}$ $ (2 +
\alpha_{\rm gas})^{-1}\,$ $(\varepsilon_{\rm th+nth} /\varepsilon_{\rm
light})$ $ (1-f_{\rm cl})$ if $\Omega_{\rm o} =0$.  Our result also
satisfies constraints from HeII absorption in the IGM (Sethi \& Nath
1997).

The true cosmological BH heating of gas in cosmological filaments
might be lower than this heat, due to the nonnegligible fraction of
nonthermal energy which does not dissipate, and which is mainly
invisible to Comptonization measurements. But this phase cools slower
than the thermal gas, due to a smaller adiabatic index.  The present
day BH energy release should therefore be slightly higher than the
thermal energy given above (present day BH released thermal and
nonthermal energy outside clusters per cosmological volume: $(2.5-
4.0)\cdot 10^{-16}\, {\rm erg\,cm^{-3}}$). This is large enough in
order to be energetically and dynamically important.  For comparison:
numerical simulation for the evolution for IGM assumes only
photoionization from quasars and get a temperature increase of several
eV ({\it e.g.}  Miralda-Escud\'e et al. 1996; and references therein).
Simulations of structure formation predict temperatures of $0.1 - 0.5$
keV within filaments, resulting from shock heating at the accretion
shocks of filaments (Kang et al.~1996)\nocite{kang96}.  This
demonstrates the need for simulations of structure formation, which
take into account the back reaction of galaxy formation induced BH
growth onto the inter-galactic medium.

\section{Conclusions} 
We estimate the contribution of energy release of black hole formation
during the quasar active epoch of the Universe for the energy budget of the
Coma cluster of galaxies and also on a cosmological scales in order to
investigate their possible energetical and dynamical influence.  Although
uncertainties are large, we find, that these energies are comparable to that
of the thermal gas of the environment. In the case of galaxy clusters a
possible hydrostatic influence of nonthermal phases was proposed as a
possible systematic effect for mass determinations of clusters, and therefore
for $\Omega_{\rm o}$ (Loeb \& Mao 1994; En{\ss}lin et al. 1997). This Letter
shows that sufficient energy seems to be injected (depending on the black
hole efficiency and black hole to spheroidal mass ratio), but the unknown
details of thermalization and spatial distribution of nonthermal phases
within the intra-cluster medium do not allow a final answer to this
question. Since large amounts of magnetized plasma should have flown out of
quasars, this scenario also can help to explain the origin of the observed
intra-cluster magnetic fields. On larger scales, the black hole energy
release can influence the large-scale structure formation by the feed back
reaction during galaxy formation and black hole growth. This cosmological
estimate is quite robust against changes in the unknown parameters of black
hole energy release, since it depends only on the ratio of the thermal plus
nonthermal to light production efficiency.\\[1em]
{\it Acknowledgments.} We thank J\"org Colberg and Dongsu Ryu for
discussion about scaling of the density in filaments, J\"org Rachen
for comments on the manuscript, and the anonymous referee for bringing
some recent work to our attention. TAE thanks for support by the {\it
Studienstiftung}.

\end{document}